\documentclass[lettersize,journal]{IEEEtran}
\usepackage{hyperref}
\usepackage{url}
\usepackage{amsmath,graphicx,bm,bbm,algorithm,color} 
\usepackage{amsmath, mathtools, amsfonts, bm, amssymb, amsthm,graphicx, caption, subcaption,multirow}%
\usepackage{algpseudocode,matlab-prettifier}
\def\BibTeX{{\rm B\kern-.05em{\sc i\kern-.025em b}\kern-.08em
    T\kern-.1667em\lower.7ex\hbox{E}\kern-.125emX}}
\usepackage{balance}

\begin{document}

\newcommand{\SE}{\mathrm{SD}}
\newcommand{\SEF}{\SE}
\newcommand{\sA}{\mathcal{A}}
\newcommand{\x}{\bm{x}}
\renewcommand{\c}{\bm{c}}
\renewcommand{\u}{\bm{u}}
\newcommand{\g}{\bm{g}}
\newcommand{\G}{\bm{G}}
\newcommand{\X}{{\bm{X}}}
\newcommand{\C}{{\bm{C}}}
\newcommand{\s}{\bm{s}}
\renewcommand{\S}{\mathcal{S}}
\newcommand{\vv}{\bm{v}}
\renewcommand{\c}{c}
\newcommand{\E}{\bm{E}}
\newcommand{\e}{\bm{e}}
\newcommand{\y}{\bm{y}}
\newcommand{\w}{\bm{w}}
\newcommand{\U}{{\bm{U}}}
\newcommand{\R}{{\bm{R}}}
\newcommand{\M}{\bm{M}}
\renewcommand{\H}{\bm{H}}
\newcommand{\calG}{\mathcal{G}}

\newcommand{\I}{\bm{\mathcal{I}}}
\newcommand{\Y}{\bm{Y}}
\newcommand{\Zhat}{{\bm{\hat{Z}}}}
\newcommand{\Z}{{\bm{{Z}}}}
\newcommand{\D}{{\bm{D}}}
\newcommand{\DD}{{\bm{F}}}
\newcommand{\Q}{{\bm{Q}}}
\newcommand{\F}{{\bm{F}}}
\newcommand{\iidsim}{\stackrel{\mathrm{iid}}{\thicksim }}
\newcommand{\n}{{\cal{N}}}
\newcommand{\indepsim}{\stackrel{\mathrm{indep.}}{\thicksim }}
\newcommand{\dist}{\mathrm{dist}}
\newcommand{\tta}{\ta^{\text{trunc}}}
\newcommand{\A}{\bm{A}}
\newcommand{\full}{{\mathrm{full}}}

\newcommand{\sg}{{\scalebox{.6}{{(g)}}}}
\newcommand{\sgp}{{\scalebox{.6}{{(g')}}}}
\newcommand{\sone}{{\scalebox{.6}{{(1)}}}}
\newcommand{\stwo}{{\scalebox{.6}{{(2)}}}}
\newcommand{\sL}{{\scalebox{.6}{{(L)}}}}
\newcommand{\con}{\mathrm{con}}
\newcommand{\epsfin}{\eps_{fin}}

\newcommand{\sj}{{\scalebox{.6}{{(j)}}}}
\newcommand{\tildeb}{\tilde{b}}
\newcommand{\gradU}{\mathrm{GradU}} 
\newcommand{\hatgradU}{\widehat{\mathrm{GradU}}}
\newcommand{\conserr}{\mathrm{ConsErr}}
\newcommand{\err}{\mathrm{Err}}
\newcommand{\Uerr}{\mathrm{UErr}}

\newcommand{\Span}{\mathrm{span}}
\newcommand{\rank}{\mathrm{rank}}
\newcommand{\evdeq}{\overset{\mathrm{EVD}}=} 
\newcommand{\svdeq}{\overset{\mathrm{SVD}}=} 
\newcommand{\qreq}{\overset{\mathrm{QR}}=} 
\newcommand{\bi}{\begin{itemize}} \newcommand{\ei}{\end{itemize}}
\newcommand{\ben}{\begin{enumerate}} \newcommand{\een}{\end{enumerate}}
\newcommand{\vsm}{\vspace{-0.1in}}

\renewcommand{\implies}{\Rightarrow}


\newcommand{\cblue}{\color{blue}}
\newcommand{\cbl}{\color{black}}
\newcommand{\cred} {\color{red}}
\newcommand{\skipit}{ }

\newcommand{\trace}{\mathrm{trace}}

\newcommand{\qfull}{q_\full}
\newcommand{\sub}{{\mathrm{sub}}}

\newcommand{\bbf}{\mathbb{F}}
\newcommand{\dsW}{\mathds{W}}
\newcommand{\dsZ}{\mathds{Z}}

\newcommand{\mtx}[1]{\mathbf{#1}}
\newcommand{\vct}[1]{\mathbf{#1}}
\newcommand{\abs}[1]{\left|#1\right|}
\newcommand{\p}{\bm{p}}
\renewcommand{\j}{\bm{j}}
\newcommand{\uu}{{\bm{u}}}
\newcommand{\tot}{\mathrm{tot}}
\renewcommand{\a}{\bm{a}}
\newcommand{\ta}{\bm{\tilde{a}}}
\newcommand{\h}{\bm{h}}
\renewcommand{\b}{\bm{b}}
\renewcommand{\d}{\bm{d}}
\newcommand{\B}{\bm{B}}
\newcommand{\V}{{\bm{V}}}
\newcommand{\W}{\bm{W}}
\renewcommand{\s}{\bm{s}}
\renewcommand{\aa}{\bm{a}}

\newcommand{\bP}{{\bm{P}}}

\newcommand{\z}{\bm{z}}
\newcommand{\zstar}{\z^\star}
\newcommand{\indic}{\mathbbm{1}}
\newcommand{\one}{\bm{1}}
\newcommand{\ps}{\small\textbf{ProjGD-s}}

\renewcommand{\C}{\bm{C}}
\newcommand{\Chat}{\bm{\hat{C}}}
\newcommand{\cb}{\bm{c}}

\newcommand{\bz}{\boldsymbol{z}}

\newcommand{\tc}{\tilde{c}}
\newcommand{\tC}{\tilde{C}}

\setlength{\arraycolsep}{0.01cm}

\newcommand{\Bstar}{{\B^\star}}   
\newcommand{\bstar}{\b^\star}             
\newcommand{\dstar}{\d^\star}             
\newcommand{\estar}{\e^\star}
\newcommand{\Estar}{\bm{E}^\star}
\newcommand{\tb}{\rho}
\newcommand{\zbarstar}{\zbar^\star}

\newcommand{\Vcheck}{\check{\V}}
\newcommand{\Bcheck}{\check{\V}}
\newcommand{\bcheck}{\check{\v}}

\newcommand{\xhat}{\hat\x}
\newcommand{\Bhat}{\hat\B}
\newcommand{\Xhat}{\hat\X}

\newcommand{\bhat}{\hat\b}
\newcommand{\Uhat}{\hat\U}

\newcommand{\Utilde}{\widetilde\U}
\newcommand{\td}{\tilde{\bm{d}}^\star}
\newcommand{\init}{{\mathrm{init}}}

\newcommand{\Ustar}{\U^\star{}}
\newcommand{\Xstar}{\X^\star{}}
\newcommand{\xstar}{\x^\star}
\newcommand{\sstar}{\s^\star}
\renewcommand{\S}{\bm{S}}
\newcommand{\Sstar}{\S^\star}
\newcommand{\Vstar}{\V^\star{}}

\newcommand{\deltinit}{\delta_\init}
\newcommand{\deltapt}{\delta_{t}}
\newcommand{\deltaptplus}{\delta_{t+1}}

\newcommand{\bSigma}{{\bm\Sigma^*}}
\newcommand{\tSigma}{\bm{E}_{det}}
\newcommand{\sigmin}{{\sigma_{\min}^\star}}
\newcommand{\sigmax}{{\sigma_{\max}^\star}}

\newcommand{\ik}{{ki}}
\newcommand{\J}{\mathcal{J}}

\renewcommand{\P}{\bm{P}}
\newcommand{\proj}{\mathcal{P}}
\newcommand{\norm}[1]{\left\|#1\right\|}

\renewcommand{\P}{\bm{U}}
\newcommand{\Phat}{\hat\P} 
\newcommand{\Lam}{\bm\Lambda} 
\renewcommand{\L}{\bm{L}}
\newcommand{\Lstar}{\L^*}
\renewcommand{\V}{\bm{V}}

\renewcommand{\l}{{\bm{\ell}}}
\newcommand{\lstar}{\l^*}
\renewcommand{\v}{\bm{v}}
\newcommand{\tty}{\tilde\y}

\newcommand{\lhat}{\hat\l}

\newcommand{\at}{\a_t}
\newcommand{\yt}{\y_t}
\newcommand{\lt}{\l_t}
\newcommand{\xt}{\x_t}
\newcommand{\vt}{\v_t}
\newcommand{\et}{\e_t}

\newcommand{\bea}{\begin{eqnarray}}
\newcommand{\eea}{\end{eqnarray}}

\newcommand{\nn}{\nonumber}
\newcommand{\ds}{\displaystyle}

\newtheorem{theorem}{Theorem}[section]
\newtheorem{prop}[theorem]{Proposition}
\newtheorem{lemma}[theorem]{Lemma}
\newtheorem{claim}[theorem]{Claim}
\newtheorem{assu}[theorem]{Assumption}
\newtheorem{corollary}[theorem]{Corollary}
\newtheorem{fact}[theorem]{Fact}
\newtheorem{definition}[theorem]{Definition}
\newtheorem{remark}[theorem]{Remark}
\newtheorem{example}[theorem]{Example}
\newtheorem{sigmodel}[theorem]{Model}
\renewcommand\thetheorem{\arabic{section}.\arabic{theorem}}

\newcommand{\snr}{\text{SNR}}

\newcommand{\Section}[1]{\vspace{-0.1in} \section{#1}  \vspace{-0.05in}  } 
\newcommand{\Subsection}[1]{ \vspace{-0.1in} \subsection{#1}  \vspace{-0.01in} }   
\newcommand{\tk}{\tilde{k}}
\newcommand{\tl}{{\ell}}
\newcommand{\totl}{L}

\newcommand{\bfpara}[1]{ {\bf #1. }} 
\newcommand{\Item}{\item} 

\renewcommand{\P}{\bm{P}}
	\newcommand{\kron}{\otimes}
\newcommand{\Uvec}{{\U_{vec}}}
\newcommand{\Zvec}{{\Z_{vec}}}
\newcommand{\ym}{{\y_{(mag)}}}

\newcommand{\eps}{\epsilon}
\newcommand{\ev}{\mathcal{E}}
\newcommand{\mbar}{\bar{m}}
\newcommand{\jk}{jk}

\renewcommand{\bhat}{\b}  \renewcommand{\Bhat}{\B}
\renewcommand{\xhat}{\x}  \renewcommand{\Xhat}{\X}
\newcommand{\rhat}{\hat{r}}
\newcommand{\ty}{\tilde{\y}}
\renewcommand{\tty}{\tilde{\ty}}
\newcommand{\deltaFt}{\delta_{F,t}}
\newcommand{\deltaFzero}{\delta_{F,0}}
\newcommand{\zbar}{\bar{\z}}
\newcommand{\deltaF}{{\delta^F}}
\newcommand{\yhat}{\hat{\y}}
\newcommand{\sigmamin}{\sigma_{\min}} \newcommand{\sigmamax}{\sigma_{\max}}

\newcommand{\Zstar}{{\Z}^*}
\newcommand{\inperr}{\mathrm{InpErr}}

\newcommand{\epsconscalar}{\eps_{\con,sc}}

\newcommand{\trnc}{\mathrm{trnc}}
\newcommand{\inp}{\mathrm{in}}

\newcommand{\errph}{ \ \mathrm{ErrPh} \ }

\renewcommand{\r}{\bm{r}}
\newcommand{\mx}{\mathrm{mx}}
\title{
Low Latency and Generalizable Dynamic MRI via L+S Alternating GD and Minimization
}
\author{Silpa Babu,  \IEEEmembership{Member, IEEE}, Sajan Goud Lingala,  \IEEEmembership{Member, IEEE}, and Namrata Vaswani,  \IEEEmembership{Fellow, IEEE}
\thanks{This work was supported by NSF under Grant CCF-2341359 and NIH under Grant R01-HL173483. A part of this work will appear in ICASSP 2025 \cite{ICASSP_2025}, this short version does not do an exhaustive comparison of various methods and almost no prospective ones.
}
\thanks{Silpa Babu and Namrata Vaswani are with the Department of Electrical and Computer Engineering, Iowa State University, USA (e-mail:
sbabu@iastate.edu; namrata@iastate.edu). Sajan Goud Lingala is with the Department of Biomedical Engineering, University of Iowa, USA (e-mail: sajangoud-lingala@uiowa.edu).}}

\maketitle


\begin{abstract}
In this work, we develop novel MRI reconstruction approaches that are accurate, fast and low-latency for a large number of dynamic MRI applications, sampling schemes and sampling rates; without any problem-specific parameter tuning. We refer to this property of a single algorithm,  without parameter tuning, being accurate and fast for many settings as  ``generalizability".  
Generalizability is possible only for simple (few parameter) models such as low-rank (LR) or LR + sparse (L+S), and for simple few parameter algorithms based on these models, which is what we develop and evaluate in this work. 

Our first contribution is a novel  Alternating Gradient Descent (GD) and Minimization or AltGDmin based solution for L+S matrix recovery from column-wise undersampled measurements (the dynamic MRI problem) and simulation experiments demonstrating that this indeed converges correctly. Next, we develop a 3-level hierarchical L+S model based extension of AltGDmin-L+S and show its generalizability for a large number of dynamic MRI applications, sampling schemes and sampling rates; without any problem-specific parameter tuning.  
Our second main contribution is a novel few shot (FS) learning based modification of this approach that provides a low latency  (also know as near real-time) solution. After a short initial mini-batch delay, it is able to reconstructs each new MRI frame as soon as its MRI data arrives, and has per-frame processing time comparable to or faster than the scan time per frame.  We refer to the resulting method as AGM-L+S-FS. A simpler LR-only special case is also evaluated. 
We demonstrate the generalizability, speed and low-latency, of our proposed algorithms via experiments on 6 prospective datasets, and 12 retrospective datasets that span multiple different dynamic MRI applications (cardiac, speech, abdomen), sampling schemes (Cartesian, pseudo-radial, radial, spiral), and sampling rates. Detailed comparisons with existing state-of-the-art low latency methods, LR and L+S methods, and unsupervised deep learning methods are provided.
%
\end{abstract}


\Section{Introduction}
\label{sec:introduction}
Magnetic Resonance Imaging (MRI) is a safe and non-invasive modality used to acquire cross-sectional images of human organs.
After appropriate pre-processing, MRI scan data can be interpreted as samples in the spatial frequency domain (k-space), which are related to the underlying image via the Fourier transform. These are acquired sequentially one row, or radial spoke, of coefficients at a time, making the scanning slow. The scan can be accelerated by undersampling the k-space according to a pre-specified trajectory, and using various modeling assumptions to still obtain an accurate reconstruction. For use in interventional radiology (interactive) applications such as low-latency ungated free breathing cardiac MRI or visualizing articulatory movements in biofeedback type experiments, there is a need to obtain low-latency reconstructions. 




\Subsection{Existing Work}

Accelerated dynamic MRI algorithms have been extensively studied.  Broadly speaking there are a few classes of recent approaches -- sparsity (compressed sensing) based methods, e.g., \cite{sparsemri} and follow-up works, low-rank (LR), including Low-Rank plus Sparse (LpS), based methods \cite{zhao2012image,lingala2011accelerated,otazo2015low,lin_fessler,lrpr_gdmin_mri_jp},
supervised deep learning (DL) methods, e.g., \cite{cinevn}, more recent unsupervised DL methods \cite{varMRI,varMRI_0,deblur,blumenthal2024self}, and low-latency MRI solutions \cite{real_time1,real_time2,real_time3,vs_spiral,lassi,onair,rt_mri_review}.

Supervised DL methods require a lot of training data which is not easy to obtain for dynamic imaging (except for a few settings such as breath-held cardiac) and hence are unusable for most applications. Unsupervised DL methods do not require training data, making them usable for new datasets (in principle). However, in practice, as we demonstrate, without architecture or parameter tuning, these also do not work well for unseen data;  see Fig. \ref{retro_perf}. Additionally, they are very slow/expensive at inference time since the DL model is both learning parameters and inferring using the input k-space dataset. For the same reason, these cannot be used to get a low-latency MRI reconstruction, where fast, frame-by-frame updates are critical. %
Low-latency approaches \cite{real_time1,real_time2,real_time3,lassi,onair,vs_spiral} are more generalizable than DL methods, but do not work well for datasets with significant motion. The reason is most of these methods combine the scan data from multiple consecutive frames, and/or use the high spatial frequency information from just the first (reference) frame.

Overall LR-based (including LpS) methods are the most generalizable since these are able to handle some amount of motion, and have few tuning parameters and model type choices. 
However, most existing algorithms based on LR or LpS models are slow and memory-intensive; many attempt to solve a convex relaxation of the true problem. Also, most have not been carefully evaluated for exact recovery for simulated datasets. 

%

To enable low-latency reconstruction, we incorporate the linear few shot learning (FSL) idea from recent works \cite{du2020few,netrapalli,altgdmin_icml}. 
Linear FSL learns the model parameters for a new linear regression task using highly undersampled data for that task. The assumption is that this task is similar to an existing set of $\alpha$ similar tasks for which a common ``representation'' can be learned. In the linear case, the similarity is modeled by assuming that the matrix formed by the model parameter vectors for these tasks is LR. In existing literature \cite{mri_seg1}, FSL is only used for DL methods and means something completely different there. 

 In recent work  \cite{lrpr_gdmin_mri_jp}, we adapted the Alternating GD and minimization (AltGDmin) based algorithm for low rank column-wise matrix sensing from \cite{lrpr_gdmin}  for fast and accurate dynamic MRI reconstruction.  To do this, we introduced a 3-level hierarchical LR model on the MRI datasets. Extensive experiments showing the power and speed of this method were shown. An attempt was also made to develop a low-latency extension but its performance was not good (see the last column of Table \ref{error_rt}). The limitation of LR models is that these cannot handle sudden motion in small (sparse) parts of the image sequence, which could be due to abnormalities in human organ function, or the rapid movement of the tongue tip during speech.  For such data, an LpS model is known to be much better model \cite{otazo2015low}. We demonstrate this in Fig. \ref{Fig_vm}, \ref{batch_speech}, and Table \ref{error_rt}.  

\Subsection{Contributions}
Our first contribution is a novel  Alternating Gradient Descent (GD) and Minimization or AltGDmin based solution for LpS matrix recovery from column-wise undersampled measurements (the dynamic MRI problem) and simulation experiments demonstrating that this indeed converges correctly. See Table \ref{error_time_init} and Fig. \ref{Fig_vm}. 
 Next, we develop a 3-level hierarchical extension of AltGDmin-LpS (AGM-LpS) and demonstrate its generalizability (accuracy and speed for a large number of dynamic MRI applications, sampling schemes and sampling rates; without any problem-specific parameter tuning) via extensive experiments on many different retrospective and prospective datasets. 
Our second main contribution is a novel few shot (FS) learning based modification of AGM-LpS that, after a short initial mini-batch delay, reconstructs each new MRI frame as soon as its MRI data arrives, and has per-frame processing time comparable to or faster than the scan time per frame. Its LR-only simplification is also evaluated. 


We demonstrate that our proposed algorithms work well (are accurate and fast) without any parameter tuning on  6 prospective datasets and 12 retrospective multi-coil k-space datasets that span multiple different applications -- speech larynx imaging, ungated cardiac perfusion, cardiac cine, free breathing cardiac OCMR, abdomen; sampling schemes -- Cartesian, pseudo-radial, radial, spiral; and sampling rates (as low as 4 radial lines per frame). 
Our work also provides an exhaustive comparison of our proposed algorithms with state-of-the-art (SOTA) approaches from various classes of accelerated dynamic MRI solutions -  one SOTA unsupervised DL method - varMRI \cite{varMRI,varMRI_0} (see Fig. \ref{batch_speech}, \ref{varMRI} and Tables \ref{error_batch});  a few of the most well-known low-latency dynamic MRI solutions -- keyhole imaging, view sharing, and OnAIR \cite{real_time1, real_time2, vs_spiral, onair} (see Fig. \ref{real}, \ref{pros_real}, \ref{ocmr_real} and Tables \ref{error_rt}); and three of the best known LR, including LpS, solutions (all batch methods) -- kt-SLR, LpS-Otazo, and LpS-Lin \cite{lingala2011accelerated,otazo2015low,lin_fessler}  (see Fig. \ref{batch_speech}, \ref{batch_cardperf} and Tables \ref{error_batch}) and our own past work on the LR-only model \cite{lrpr_gdmin_mri_jp} (see Fig. \ref{batch_speech} and Table \ref{error_rt}).  
Each algorithm is compared on all the above stated datasets, without any parameter tuning. We show that our proposed algorithms has the lowest or second lowest reconstruction error for all retrospective datasets. 

\Section{Problem setting}

\subsubsection{Goal}  This work has two main goals: (i) developing fast and generalizable LR and LpS model based algorithms for dynamic MRI, and (ii) developing novel low-latency modifications of these algorithms that retain the generalizability property. 
The goal in low-latency imaging is to  recover the $k$-th image $\zstar_k$, as soon as the k-space data (Fourier measurements) for it, $\y_k$, arrive; and we need it to be fast, the per-frame reconstruction time should be equal to or lower than the per-frame scan time.  
Our approaches assume that an initial mini-batch of $\alpha$ frames is processed as a batch, but, after this, the algorithm provide low latency MRI reconstructions.






\subsubsection{Cartesian MRI measurement model}
To explain the model simply, we assume that the unknown MRI images are vectorized and arranged as columns of an $n \times q$ matrix, $\Zstar = [\zstar_1, \zstar_2, \ldots, \zstar_q]$;
thus $\zstar_k$ is the $k$-th image of the sequence vectorized and $q$ is the total number of frames in the MRI sequence. 
Let $\bm{F}$ be an $n \times n$ matrix that models the 2D discrete Fourier Transform (DFT) operation on the vectorized image; thus $\F \z$ is the vectorized 2D-DFT of vectorized image $\z$.
A broad 1-0 matrix $\H_k$ of size $m_k \times n$  models the undersampling operation: each row of it contains exactly one 1 corresponding to the DFT coefficient that is observed. Thus $\H_k$ specifies the sampling mask.
Let $\A_k = \H_k \F$. In single-coil Cartesian MRI, one needs to recover the images $\zstar_k$ from measurements
%
$
\y_k := \A_k \zstar_k = (\H_k \bm{F}) \zstar_k, \ k \in [q].
$

In case of multi-coil  Cartesian  MRI, there are multiple  receive channels (coils), each measuring a subset of Fourier coefficients of a differently weighted version of the cross-section to be imaged. Denote the number of coils by $c$.  Let $\y_{k,j}$ denote the measurements at the $j$-th coil. Then, $\y_k^\top = [\y_{k,1}^\top, \y_{k,2}^\top, \dots, \y_{k,c}^\top]$ and $\y_{k,j} = \H_k \F \D_j \zstar_k$.
where $\D_j = diag(\d_j, j=1,2,\dots,n)$ are $n \times n$ diagonal matrices with diagonal entries (entries of the vector $\d_j$) being the coil sensitivities of the $j$-th coil.
%
%

Let $m=\max_k(m_k)$. We define the $mc \times n$ matrix  $\Y= [ (\y_1)_{\mathrm{\tiny{long}}}, (\y_2)_{\mathrm{\tiny{long}}}, \cdots,  (\y_q)_{\mathrm{\tiny{long}}}]$ with $(\y_k)_{\mathrm{\tiny{long}}}$ being the vector $\y_k$ followed by $(m-m_k)c$ zeros. Similarly let $(\A_k)_{\mathrm{\tiny{long}}}$ be an $mc \times n$ matrix with $(m-m_k)c$ rows of zeros at the end. Then, the above model can also be expressed as
$
\Y = \sA(\Zstar)= [(\A_1)_{\mathrm{\tiny{long}}} (\zstar_1), (\A_2)_{\mathrm{\tiny{long}}} (\zstar_2), \dots, (\A_q)_{\mathrm{\tiny{long}}} (\zstar_q)].	
$

\subsubsection{Radial MRI} 
The above matrix-vector model is valid when the samples are available on a Cartesian grid; either they are acquired that way or they are mapped onto a Cartesian grid, e.g., in case of  pseudo-radial sampling.
When directly using true radial samples, the main ideas above, and in our algorithms given below, are still exactly the same, except that the DFT gets replaced by non-uniform FT (NUFT), e.g., \cite{jeff_nufft}.

\section{AltGDmin (AGM) for the LpS MRI model}

In this section, we first describe the basic LpS model \cite{otazo2015low,lin_fessler}, followed by a novel fast gradient descent-based algorithm. We then propose a novel three-level hierarchical LpS model and its corresponding algorithm for dynamic MRI.
\subsection{Low-Rank Plus Sparse (LpS) model \cite{otazo2015low,lin_fessler}}
\label{LR+S}
The LpS decomposition is a well-studied model in dynamic MRI  \cite{otazo2015low,lin_fessler}.
The objective is to recover an $n \times q$ matrix $\Xstar$ from undersampled measurements of the form
\[
\y_k = \A_k \xstar_k, \quad k \in [q],
\]
where $\A_k$ is a known measurement matrix of size $m \times n$ with $m \ll n$. The vectors $\xstar_k$ and $\y_k$ denote the $k$-th columns of $\Xstar$ and $\Y$, respectively. We assume that $\Xstar = \Lstar + \Sstar$,
where $\Lstar$ is a rank-$r$ matrix with $r \ll \min(n,q)$, and $\Sstar = [\sstar_1, \dots, \sstar_q]$ is a sparse matrix. 
Any rank-$r$ matrix $\Lstar$ can be factorized as $\Lstar = \Ustar\Bstar$, where $\Ustar$ is an  ${n \times r}$ matrix with orthonormal columns, and $\Bstar = [\bstar_1, \dots, \bstar_q]$ is an $r \times q$ matrix, where $\bstar_k$ denotes its $k$-th column. Thus, the $k$-th column of $\Xstar$ is given by
$
\xstar_k = \Ustar\bstar_k + \sstar_k
$.
Hence, the undersampled measurements can be written as
\[
\y_k = \A_k(\Ustar\bstar_k + \sstar_k), \quad k \in [q].
\]

\begin{assu}[Incoherence assumption: Left and Right singular vectors' are incoherent]
We assume that the left singular vectors satisfy $\max_j \|(\u^j)^*\| \le \mu \sqrt{r/n}$, where $(\u^j)^*$ denotes the $j$-th row of $\Ustar$. Similarly, each $\bstar_k$ satisfies $\max_k \|\bstar_k\| \le \mu  \sqrt{r/q} {\sigma^*_{max} }$, where $\sigma^*_{max}$ is the largest singular value of $\Lstar$.
\end{assu}
As a result, the entries of $\Lstar = \Ustar \Bstar$ are bounded as $\|\Lstar \|_{\infty} \leq \mu^2 \frac{r}{\sqrt{nq}} \sigma^*_{max}$, which implies that the magnitude of $\Lstar$ is bounded. There is no such magnitude bound is assumed for the sparse matrix $\Sstar$. In particular, this means that $\|\sstar_k\|_{\infty} \gg \| \Ustar \bstar_k\|_{\infty}$.
\subsection{Proposed algorithm: Basic AltGDmin-LpS (AGM-L+S-basic)}
\label{LRS_UBS}
We develop a fast gradient descent based algorithm to find matrices $\U$, $\B$, and $\S$ that minimizes the cost function:
\[
f(\S, \U, \B)=\sum_{k=1}^q \|\y_k-\A_k (\U\b_k+\s_k)\|^2,
\]
where $\|.\|$ denotes the (induced) $l_2$ norm of a vector/matrix.
 A standard approach to solve this problem is to use Alternating Minimization (AltMin) (actually Block Coordinate Descent (BCD) which refers AltMin over three or more subsets of variables), to iterate over alternatively updating $\S, \U,\B$ by full minimization for one keeping the other two fixed. However, these methods are slow because $\U$ appears in all $q$ terms, and hence updating $\U$ via exact minimization requires solving a very large $mq \times nr$ dimensional least squares (LS) problem each time. To address this, we replace the slow $\U$ update with a projected gradient descent step. This significantly speeds up convergence while maintaining reconstruction quality. 
 This modification is best understood as an AltGDmin type modification of the BCD algorithm. We henceforth refer to it as AltGDmin. 


\subsubsection{AGM Iterations}
Given an intial estimate of $\S, \U, \B$, the AGM algorithm alternates between the following three steps. We explain the initialization below in Sec \ref{Uinit}. 

(1) For each new estimate of $\U$ and $\B$, we update $\S$ by minimizing $f(\S, \U, \B)$ over $\S$, keeping $\U$ and $\B$ fixed. Because the measurements are column-wise decoupled for the different columns of $\S$, this simplifies to solving $q$ independent sparse recovery problems:
\[
\arg\min_{\s_k} \|(\y_k - \A_k \U \b_k) - \A_k \s_k\|^2, \ k \in [q].
\]

(2) Next, we update $\U$ using a single gradient descent (GD) step followed by QR-based orthonormalization: $\U^+ = QR(\U - \eta \nabla_U f(\S, \U, \B))$, where $\eta$ is the step-size (learning rate), and the gradient is given by
$\nabla_U f(\S, \U, \B) =\sum_k \A_k^{\top} (\y_k - \A_k ( \U \b_k +\s_k)) \b_k^\top $.

(3) Finally, we update $\B$ by minimizing the cost function $f(\S, \U, \B)$ over $\B$, keeping $\U$ and $\S$ fixed. Again, due to column-wise decoupling, this reduces to solving $q$ independent $r$-dimensional least squares (LS) problems:
$
\arg\min_{\b_k} \|\y_k - \A_k \s_k- \A_k \U \b_k \|^2, \ k \in [q].
$
The closed-form solution to each problem is
\[
 \b_k= (\A_k \U)^\dag (\y_k - \A_k \s_k), 
\]
where $\M^\dag:=(\M^\top \M)^{-1} \M^\top$.
We summarize our proposed algorithm in the function \textproc{AGM-LpS} within Algorithm \ref{lplussMRI}.


\subsubsection{Initialization}
%
 \label{Uinit}

Given $\{\y_k,\A_k\}$ for all $k \in [q]$, we estimate the initial sparse matrix $\S_\init$ by solving $q$ independent sparse recovery problems:
\[
\s_k = \arg\min_{\s_k} \|\y_k - \A_k \s_k\|^2, \ k \in [q],
\]
where $\s_k$ denotes the $k$-th column of $\S_\init$.
We then compute residuals $\tilde{\y}_k := \y_k - \A_k \s_k$ and 
obtain the initialization matrix $\X_0$ as:
\begin{equation}
\X_{0} =     \left[ \A_1^\top \ty_{1},\A_2^\top \ty_{2},  ..., \right.  \\
 \left. \A_q^\top \ty_{q} \right]
 \label{X0}
\end{equation}
We compute the initial subspace estimate $\U_\init$ as the top $r$ left singular vectors of $\X_0$ with $r$ estimated as follows \cite{lrpr_gdmin_mri_jp}
\begin{equation}
r = \min \{\hat{r}: \sum_{j=1}^{\hat{r}} \sigma_j^2(\X_0) \ge 0.85 \sum_{j=1}^{r_{big}} \sigma_j^2(\X_0) \},
\label{r_estim}
\end{equation}
where $r_{big} = \min(n, q, \min_k m_k)/10$ and $\sigma_j$ is the $j$-th singular value of $\X_0$. Then, we compute each column of the initial estimate $\B_\init$ as $
 \b_k= (\A_k \U)^\dag (\y_k - \A_k \s_k)
$. We summarize the spectral initialization in Algorithm \ref{lplussMRI} within the function \textproc{AGM-LpS-INIT}. 
\subsubsection{Parameter setting} 
The rank $r$ is set as described earlier. The step size is chosen as $\eta = 0.14 / \|\nabla_U f(\S_\init, \U_\init, \B_\init)\|$. In the simulated studies, we used hard thresholding to ensure convergence, while in the MRI experiments, we used soft thresholding to promote temporal smoothness.

During initialization, the threshold parameter is set to a relatively large value of $0.07 \|\C\|\infty$, where each column of $\C$ is computed as $\c_k = \A_k^\top \y_k$. This large threshold ensures that the low-rank component is not mistakenly captured in the initial sparse estimate. In subsequent iterations, the threshold is reduced to $0.04 \|\C\|_\infty$, where $\c_k = \A_k^\top (\y_k - \A_k \l_k)$, to more accurately recover the sparse component after the low-rank structure has been estimated.

In the simulated study, we retained the largest $\rho$ entries (in magnitude) of $\C$ during thresholding. Alternatively, the same soft thresholding strategy described above can also be used.
Here, $\| \cdot \|\infty$ denotes the maximum absolute entry of a matrix or vector.

The algorithm runs for a maximum of $T_{\max}$ iterations but may terminate early if the estimates do not change much over two consecutive iterations; specifically, the algorithm stops if $\text{NMSE}(\X_t,\X_{t-1})<0.09$ and $ \text{NMSE}(\X_{t-1},\X_{t-2})< 0.09$.

\subsection{Proposed 3-level hierarchical LpS model for dynamic MRI} \label{approx_lrscs_model}
\label{LR+S_mri}
We modify our proposed AGM-LpS algorithm to better handle real-world image sequences by introducing two additional steps, similar to the approach in \cite{lrpr_gdmin_mri_jp}. In this work, we focus specifically on dynamic MRI data.
Most single slice MRI sequences contain a baseline component that remains approximately constant across frames, which we denote as the “mean” image. 
Secondly, even after subtracting this mean, MR image sequences are only approximately low-rank; that is, the residual obtained after removing the mean and low-rank components can still be large. To better capture this structure, we introduce an additional sparse component, resulting in a LpS model for the mean-subtracted data. The residual remaining after removing the mean, low-rank plus sparse components is significantly smaller in magnitude and thus easier to estimate. Likewise, the low-rank and sparse components can be more accurately estimated after removing the estimated mean.
Thus the following model is the most appropriate for dynamic MRI: the $k$-th vectorized MR image, $\zstar_k$, satisfies
\begin{equation}
\label{mri_ls}
\zstar_k = \zbarstar + \Ustar\bstar_k + \sstar_k + \estar_k , \ \forall \ k \in [q],
\end{equation}
where $\zbarstar$ is the mean image, $\Ustar\bstar_k$ is the $k$-th column of  the rank-$r$ matrix $\Lstar$, with $r \ll \min(n, q)$, $\sstar_k$ is the $k$-th column of the sparse matrix $\Sstar$, and $\estar_k$ is the $k$-th column of $\Estar$, denotes the unstructured component.

This model can be interpreted as a 3-level hierarchical LpS model: the first level corresponds to the rank-1 ``mean image'' matrix, $\zbarstar \bm{1}^\top$, where $\bm{1} \in \mathbb{R}^q$ is the all-ones vector;
the second level consists of low rank  plus sparse matrix, $\Lstar + \Sstar$, where $\Lstar$ models slowly changing components across frames and  and $\Sstar$ models sparse, large outliers;
and the third level consists of the small magnitude, full-rank residual matrix $\Estar$, with rank $\min(n, q)$.

\subsection{Proposed algorithm: AGM-LpS for Dynamic MRI}
\label{3L_LS}

\begin{algorithm}[t]
\caption{\sl{AltGDmin-LpS for Dynamic MRI (AGM-LpS)}}
\label{lplussMRI}
\begin{algorithmic}[1]
\small
   \State {\bfseries Input:} $\{\y_k, \A_k, k \in [q]\}$
   \State Set $\tau =50$
 \State  $\y = [\y_1^\top,\ldots,\y_q^\top]^\top$, $\A = [\A_1^\top,\ldots,\A_q^\top]^\top$
\State $\zbar \leftarrow \Call{CGLS}{\y, \A, 0,10}$
\State $\ty_k := \y_k-\A_k \zbar, \quad \forall k \in [q]$
\State $[\S_\init, \U_\init] \leftarrow \Call{AGM-LpS-INIT}
{  \{\ty_k, \A_k, k \in [q]\}}$
\State $[\S, \U, \B] \leftarrow \Call{AGM-LpS-It} { \{\ty_k, \A_k, k \in [q]\}, \S_\init, \U_\init, \tau}$
\State $\tty_k := \ty_k - \A_k\U\b_k - \A_k\s_k, \quad \forall k \in [q]$
\State $\e_k \leftarrow \Call{CGLS}{\tty_k, \A_k,0,3}, \quad \forall k \in [q]$
\State      {\bfseries Output:} $\Z:= [\z_1,\ldots,\z_q]$, with $\z_k= \zbar+\U \b_k + \s_k + \e_k $.

\Statex \Function{AGM-LpS-INIT}{$\{ \y_{k}, \A_k, k \in [q]\}$}
\State $\s_k =$ Thresh$(\A_k^\top\y_k), \quad \forall k \in [q]$
\State $\yhat_k := \y_k-\A_k\s_k, \quad \forall  k \in [q]$
\State  $\X_0 := [\A_1^\top  \yhat_1,\A_2^\top  \yhat_2,  \cdots, \A_q^\top  \yhat_q]$.

\State  Set $\r$ as the smallest integer for which
$$\sum_{j=1}^r \sigma_j(\X_0)^2 \ge (b/100) \cdot \sum_{j=1}^{\min(n,q)/10} \sigma_j(\X_0)^2, \ b=85. \quad (**)$$

\State  Set $\U \leftarrow $ top $r$ left singular vectors of $\X_0$
\State      \Return {$\S, \U$}
\EndFunction
\Statex \Function{AGM-LpS-It}{$\{\y_{k}, \A_k, k \in [q]\}, \S_\init,\U_\init, \tau$}
\State $\S \leftarrow \S_\init$
\State $\U \leftarrow \U_\init$

\For{$t=1$ {\bfseries to} $\tau$}
\State $\bhat_{k} = (\A_{k}  \U)^\dagger (\y_{k}-\A_k\s_k),  \quad \forall k \in [q]$
\State $\s_k =$Thresh($ \A_k^\top(\y_k -\A_k\U\b_k)$)$, \quad \forall k \in [q]$
  \State $\nabla_U f  = \sum_{k=1}^q  \A_{k}^\top (\A_{k} (\U \bhat_{k}+\s_k) - \y_{k}) \bhat_{k}^\top$

  \State If $t=1$, set $\eta = 0.14/\|\nabla_U f \|$.

  \State   $\ds \Uhat^+   \leftarrow \U - \eta \nabla_U f$; $\Uhat^+ \qreq \U^+ \R^+$.

\State Set $\U \leftarrow \U^+$
\EndFor
\State      \Return {$\S, \U,\B$}
\EndFunction
\end{algorithmic}
\begin{algorithmic}[1]
\Statex \Function{CGLS}{$\y_{k}, \A_k, \w_\init, \text{max-iter}$}
\State   solves $\arg\min_{\w} \|\y_k-\A_k \w\|^2$ using conjugate gradient with initialization $\w_{init}$.
\\ \% (function in MATLAB and Python)
\State      \Return {$\w$}
\EndFunction

\end{algorithmic}

\noindent {\scriptsize ** $\sigma_j(\X_0)$ denotes the $j$-th singular value of $\X_0$.}
\end{algorithm}


We develop a 3-level hierarchical algorithm that first recovers $\zbar^*$, then $(\U\b^*_k + \s_k^*)$'s, and finally $\e^*_k$'s. Under the modeling assumption: $\| \zbar^*\| \gg \|\U\b^*_k + \s_k^*\| \gg \|\e^*_k \|$, the recovery of $\zbar^*$ becomes: 
\[
\min_{\tilde\zbar} \sum_{k=1}^q \|\y_k - \A_k \tilde\zbar\|^2.
\label{minx}
\]
\label{agm-mri-ls}
Although this least squares problem has a closed-form solution, directly computing it is memory-intensive due to the inversion of an $n \times n$ matrix, where $n$ is the number of pixels in a single image. To avoid this, we use the Conjugate Gradient for Least Squares (CGLS) algorithm, implemented in MATLAB or Python. Define $\y = [\y_k^\top, k \in [q]]^\top$ and $\A = [\A_k^\top, k \in [q]]^\top$.
We call $\mathrm{CGLS}( \y, \A, \zbar_\init=\bm{0}, \text{max-iter}=10)$ and denote its output by $\zbar$.
Next, we compute the residuals  $\ty_k := \y_k -\A_k \zbar$ for all $k \in [q]$, and then call the function AGM-LpS-INIT, followed by AGM-LpS in Algorithm~\ref{lplussMRI}, to estimate $\S$, $\U$ and $\B$ using the inputs ${\ty_k}$'s and ${\A_k}$'s.
Then,  for each $k \in [q]$,  we estimate $\e_k^*$ by running three iterations of the Conjugate Gradient for Least Squares (CGLS) algorithm to solve
\[
\min_{\e} \|\tilde{\ty}_k- \A_k \e \|^2, 
\]
where
$\tilde{\ty}_k:= \y_k - \A_k(\zbar + \U\b_k + \s_k),
$
and denote its output by $\e_k$.

Finally, each reconstructed vectorized MRI frame is given by $\z_k := \zbar + \U \b_k + \s_k + \e_k$, and the matrix of reconstructed frames is $\Z := [\z_1, \z_2, \dots, \z_q]$.
The overall algorithm is summarized in Algorithm~\ref{lplussMRI}.

\section{Few Shot AGM-LpS for Low Latency MRI}
To enable low-latency reconstruction, we introduce a slow-changing, mini-batch-based version of the model. We first describe the slow-changing three-level hierarchical LR model, where the $\S$ is set to zero (Sec \ref{MR_model}), followed by the slow-changing three-level hierarchical LpS model (Sec \ref{MR_model2}). The corresponding algorithms are detailed in Sec \ref{fsmri} and \ref{fsmri_ls}, respectively.
\Subsection{Proposed Slow-Changing 3-Level Low-Rank (LR) Model} 
\label{MR_model}
We consider a simplified case where the sparse component $\Sstar$ is set to zero. Thus, (\ref{mri_ls}) simplifies to $
\zstar_k = \zbarstar + \Ustar\bstar_k + \estar_k , \ \forall \ k \in [q]
$.
We introduce the following slow-changing hierarchical 3-level LR model on the image sequence.
Split the $q$-frame sequence into $q/\alpha$ mini-batches, each of size $\alpha$ frames. Assume here that $q/\alpha$ is an integer. Define the mini-batches, $\I_\ell:=[(\ell-1)\alpha+1, \ell \alpha]$ and let $\Zstar_{(\ell)}:= [\zstar_k, k \in \I_\ell]$.
We assume that 1) each image within the mini-batch is a sum of three layers -- a common (rank $r=1$) component; a LR (rank $r$) component that models the fact that image changes over time depend on a much smaller number, $r$, of factors than the sequence length $\alpha$; and a full-rank (rank $\min(n,\alpha)$) component with no temporal dependencies. Also, 2) the magnitude of each layer is much smaller than that of the previous layer.  3) Over consecutive mini-batches, the ``mean'' image changes slowly;  and the same is true for  the column-span of the LR component.  This last assumption is the one that helps us to obtain a near real-time  algorithm.

 To define the model mathematically, we first introduce some notation. Let $\mathbf{1}$ denote a vector of $\alpha$ ones, $\mathbf{I}$ denote the identity matrix, and $\|.\|_F$ denotes the Frobenius norm;
for two $n \times r$ matrices with orthonormal columns, $\U_1, \U_2$,  $\SEF(\U_1, \U_2) := \|(\mathbf{I} - \U_1 \U_1^\top) \U_2\|_F$ denotes the Subspace Distance (SD) between the subspaces spanned by their columns. 
The SD between two $r$-dimensional subspaces is a real number between 0 and $\sqrt{r}$.
Then, for all $\ell =1,2,\dots, q/\alpha$,
\begin{align}
& \Zstar_{(\ell)}: =  \zbarstar_{(\ell)} \one^\top  +  \Ustar_{(\ell)} \Bstar_{(\ell)} + \Estar_{(\ell)}, \notag \\
&  \| \Estar_{(\ell)} \|_F  \ll \| \Bstar_{(\ell)}\|_F  \ll  \|\zbarstar_{(\ell)} \| \sqrt{\alpha},  \notag \\
& \|\zbarstar_{(\ell)} - \zbarstar_{(\ell-1)}\|  \ll \|\zbarstar_{(\ell-1)}\|,  \notag \\
& \SEF(\Ustar_{(\ell)}, \Ustar_{(\ell-1)}) \ll \sqrt{r}
\label{mod}
\end{align}
with $\zbarstar_{(0)} = \bm{0}_{n}$ and $\Ustar_{(0)} = \bm{0}_{n \times r}$. In the above model, $\zbarstar$ is the common ``mean image'' component for the entire mini-batch, $\Ustar$ is a matrix with orthonormal columns whose span equals the column-span of the LR component, $\Bstar = [\bstar_k, k \in \I_\ell]$ with each $\bstar_k$ being an $r$ length vector, and $\Estar =  [\estar_k, k \in \I_\ell]$ is the full-rank component with no model assumed. This models the finer details that are different in each image.

\Subsection{Proposed algorithm: Dynamic Few Shot Low-Rank MRI (FewShot-LR-MRI)} \label{fsmri}

%
%

We develop dynamic few shot low rank MRI using altGDmin (AGM) (FS-LR-MRI)  as follows. 

\subsubsection{Initial mini-batch computation}
We wait for the undersampled k-space data of the first mini-batch of \(\alpha\) frames, $\y_k$, $k \in \I_{1}$, and estimate $\zbar_{(1)}$ by solving the least-squares problem: $\arg\min_{\zbar} \sum_{k \in \I_1} \|\y_k - \A_k \zbar\|_2^2$.  As discussed earlier (see Sec \ref{agm-mri-ls}), we use the Conjugate Gradient for Least Squares (CGLS) method:  $\mathrm{CGLS}( \y_{(1)}, \A_{(1)}, \zbar_\init=\bm{0}, \text{max-iter}=10)$, where $\y_{(1)} = [\y_k^\top, k \in \I_1]^\top$ and $\A_{(1)} = [\A_k^\top, k \in \I_1]^\top$.
%
Next, we compute the residual measurements as, $\ty_k = \y_k - \A_k \zbar_{(1)}$, $k \in \I_{1}$. 
We then run $\Call{AGM-LR-INIT}{ \{ \ty_k, \A_k, k \in \I_{1} \}}$ to estimate the initial subspace $\U_\init$. 
Next, we run the AGM-LR sub-routine $\Call{AGM-LR}{ \{ \ty_k, \A_k, k \in \I_{1} \}, \U_\init, \tau=50}$, whose output is the final estimate of $\U_{(1)}$.

\subsubsection{low-latency algorithm}
Starting at time $k= \alpha+1$, we use
the slow mean change and slow subspace change assumptions to obtain low-latency estimates as follows. 

For all $k \in \I_\ell$, set $\zbar = \zbar_{(\ell-1)}$ and $\U = \U_{(\ell-1)}$. For $\ell=2$, we obtain these from the initialization explained above. For $\ell > 2$, these are updated as explained below.
Compute $\ty_k = (\y_k - \A_k \zbar)$ and use this to obtain
\[
\b_k = (\A_k \U)^\dag \ty_k,
\]
and
\[
\e_k = \arg\min_{\e} \|(\ty_k - \A_k \U \b_k) - \A_k \e\|^2 \text{ s.t. } \|\e\|^2 \le \lambda  
\]
In the above, the constraint enforces the assumption that $\e_k$ has small magnitude. In practice, this can be approximately solved quickly by starting with a zero initialization and using only a few CGLS iterations for the unconstrained problem. Thus, we obtain $\e_k = \mathrm{CGLS}(\ty_k - \A_k \U \b_k, \A_k,\e_\init=\bm{0}, \text{max-iter}=3)$.
Finally, output the low-latency estimate as $\z_k = \zbar + \U \b_k + \e_k$

\subsubsection{Mini-batch updates of the initial estimates of $\bar{z},U$}
If we have limited computational power, we would compute $\zbar, \U$ only once using the first mini-batch. However, for image sequences with significant motion, this does not work well \cite{lrpr_gdmin_mri_jp}. The reason is both $\zbar$ and $\U$ do change over time, albeit slowly. Instead, a better idea is to also update $\zbar, \U$ every $\alpha$ frames in a mini-batch fashion, using the slowly changing assumption. This assumption makes the updating quick while keeping the algorithm accuracy comparable or better than that of a batch method. The accuracy is often better because slow changing model is a better one than a fixed subspace model.
Slow $\zbar$ change means that we now need to solve
\[
\arg\min_{\zbar} \sum_{k \in \I_\ell} \|\y_k - \A_k \zbar\|_2^2  \text{ s.t. }  \|\zbar - \zbar_{(\ell-1)}\|^2 \le \lambda  
\]
We can get a quick approximate solution for this by using CGLS for the unconstrained problem with using the previous $\zbar$ as the initialization, that is, $\zbar_{(\ell-1)}$, and using only 2 iterations instead of 10, i.e., use
$\mathrm{CGLS}(\y_{(\ell)}, \A_{(\ell)}, \zbar_{(\ell-1)}, \text{max-iter}=2)$ with  $\y_{(\ell)} = [\y_k^\top, k \in \I_\ell]^\top$ and $\A_{(\ell)} = [\A_k^\top, k \in \I_\ell]^\top$.
Slow subspace change means that we do not need the time-consuming spectral initialization step, and that we can use much fewer AGM-LR iterations. We use $\mathrm{AGM-LR}( \{ \ty_k, \A_k, k \in \I_{\ell} \}, \U_{(\ell-1)}, \tau=15)$. Here  $\ty_k = \y_k - \A_k \zbar_{(\ell)}$. Use its output as the final $\U_{(\ell)}$.

This algorithm for updating $\zbar$ and $\U$ is also computing $\B$. By also adding one more quick step that updates $\e_k$, it gives us an improved delayed reconstruction of the image sequence.
We summarize this in Algorithm \ref{fs_mri_lr}. 

Our method can be interpreted as Dynamic Few Shot Learning. The mini-batch updates help dynamically update the linear representations, $\zbar$ and $\U$ (really its column span), every so often, while the real-time component uses these for few shot learning of the images. Here the images are the model parameters.
\begin{algorithm}
\caption{\sl{AGM-LpS-FS (LR only)}}
\label{fs_mri_lr}
\begin{algorithmic}[1]
\small
\State Wait for first $\alpha$ frames; then we have $\{\y_k$, $\A_k$, $k \in \I_1\}$.
\State $[\zbar_{(1)}, \U_{(1)}, \Z_{(1)}] \leftarrow$ \Call{Delay}{$\{\y_k, \A_k, k \in \I_{(1)}\},  \ell = 1$}
\For {$k= \alpha+1$ {\bfseries to} $q/\alpha$}
\State $\ell = \lceil k / \alpha \rceil $
\State $ \z_k \leftarrow$ \Call{low-latency}{$\y_k, \A_k, \ell$}
\If {$ k \bmod \alpha =0$} 
\State $[\zbar_{(\ell)}, \U_{(\ell)}, \Z_{(\ell)}] \leftarrow$ \Call{Delay}{$\{\y_k, \A_k, k \in \I_\ell\}, \ell$}
\EndIf
\EndFor
\Statex \Function{delay}{\{$\y_{k}, \A_k, k \in \I_\ell$\}, $\ell$}
\State $\y_{(\ell)} = [\y_k^\top, k \in \I_\ell]^\top$, $\A_{(\ell)} = [\A_k^\top, k \in \I_\ell]^\top$
\If{$\ell = 1$}
\State $\zbar \leftarrow $\Call{ CGLS}{ $\y_{(\ell)}, \A_{(\ell)}, \bm{0}, 10$}
\State $\ty_{k} := \y_{k} - \A_{k}\zbar, \quad \forall k \in \I_1$
\State $ \U_\init \leftarrow \Call{AGM-LR-INIT}
{  \{\ty_k, \A_k, \I_\ell\}}$
\State $\tau = 50$
\Else
\State $\zbar \leftarrow  $\Call{ CGLS}{$\y_{(\ell)}, \A_{(\ell)}, \zbar_{(\ell-1)},2$}
\State $\ty_{k} \leftarrow \y_{k} - \A_{k}\zbar$, $\quad \forall k \in \I_\ell$
\State $\U_\init \leftarrow \U_{(\ell-1)}$
\State $\tau = 15$
\EndIf  
\State $[\U, \B] \leftarrow$ \Call{AGM-LR}{$ \{ \ty_k, \A_k, k \in \I_{\ell} \}, \U_\init, \tau$}
\State $\e_k \leftarrow $\Call{ CGLS}{$\ty_k - \A_k \U \b_k, \A_k, \bm{0},3$}, $\quad \forall k \in \I_\ell$
\State      {\bfseries Output:} $\zbar_{(\ell)} \leftarrow \zbar$, $\U_{(\ell)} \leftarrow \U$
\State  {\bfseries Output:} Delayed recons $\z_{k} =\zbar  +\U  \bhat_{k} +  \e_{k}$,  $ \forall k \in \I_\ell$
\EndFunction
\Statex \Function{low-latency}{$\y_{k}, \A_k, \ell$}
\State Get $\U \leftarrow \U_{(\ell-1)}$ and $\zbar \leftarrow \zbar_{(\ell-1)}$
\State  $\ty_{k} := \y_{k} - \A_{k} \zbar$
\State $\bhat_{k} =(\A_{k} \U)^\dagger \ty_{k}$
\State  $\tty_{k} := \ty_{k} - \A_{k} \U\b_k $
\State $\e_k \leftarrow \Call{CGLS}{\tty_k, \A_k, \bm{0},3}$
\State $\z_{k} = \zbar +\U\b_k + \e_k $
\State {\bfseries Output:} $\z_{k} $
\EndFunction

\Statex \Function{AGM-LR-INIT}{\{$\y_{k}, \A_k, k \in [q]$\}}
\State Compute $\X_0 := [\A_1^\top \y_1, \A_2^\top \y_2, \cdots, \A_q^\top \y_q]$.

\State  Set $\r$ as the smallest integer for which
$$\sum_{j=1}^r \sigma_j(\X_0)^2 \ge (b/100) \cdot \sum_{j=1}^{\min(n,q)/10} \sigma_j(\X_0)^2, \ b=85. \quad (**)$$
\State  Set $\U \leftarrow $ top $r$ left singular vectors of $\X_0$
\State      \Return {$ \U$ }
\EndFunction

\Statex \Function{AGM-LR}{$\{\y_{k}, \A_k, k \in [q]\}$, $\U_\init$, $\tau$}
\State $\U \leftarrow \U_\init$
\For{$t=1$ {\bfseries to} $\tau$}
 \State $\bhat_{k} =  (\A_{k}  \U)^\dagger \y_{k} , \quad \forall k \in [q]$

  \State $\nabla_U f (\U, \Bhat)= \sum_{k =1}^q \A_{k}^\top (\A_{k} \U \bhat_{k} - \y_{k}) \bhat_{k}^\top$

  \State If $t=1$, set $\eta = 0.14/\|\nabla_U f (\U, \Bhat)\|$.

  \State   $\ds \Uhat^+   \leftarrow \U - \eta \nabla_U f(\U, \Bhat)$; $\Uhat^+ \qreq \U^+ \R^+$.

\State Set $\U \leftarrow \U^+$
\EndFor
\State      \Return {$\U,\B$ with $\B=[\bhat_{1}, \bhat_{2},\dots, \bhat_{q}]$}
\EndFunction
\end{algorithmic}
{\scriptsize ** $\sigma_j$ be the $j$-th singular value of $\X_0$. }
\end{algorithm}
\Subsection{Proposed model and algorithm: slow-changing 3-level LpS model and FS-LpS-MRI}
\label{MR_model2}
We extend the slow-changing hierarchical 3-level LR model described earlier by including a sparse component to better capture abnormalities or localized changes. Specifically, for each mini-batch $\ell$, the dynamic image sequence $\Zstar_{(\ell)}$ is modeled as:
\[
\Zstar_{(\ell)} = \zbarstar_{(\ell)} \one^\top + \Ustar_{(\ell)} \Bstar_{(\ell)} + \Sstar_{(\ell)} + \Estar_{(\ell)}.
\]
Under the modeling assumption:
\[
  \| \Estar_{(\ell)} \|_F \ll  \| \Ustar_{(\ell)}\Bstar_{(\ell)}+ \Sstar_{(\ell)}\|_F   \ll \sqrt{\alpha}\|\zbarstar_{(\ell)} \|, \quad \ell \in [q/\alpha],
\label{mod}
\]
all other assumptions including the slow variation of the mean image and the subspace, remain unchanged.

\subsubsection{Algorithm}
 \label{fsmri_ls}
 We extend the FS-LR-MRI framework to handle the Low-Rank plus Sparse (LpS) decomposition of dynamic MRI sequences.\\
Mini-batch updates:
For the first mini-batch of frames $\I_1$, we estimate the mean image $\zbar_{(1)}$ by solving
\[
\arg\min_{\zbar} \sum_{k \in \I_1} \|\y_k - \A_k \zbar\|_2^2,
\]
 using conjugate gradient least squares (CGLS). We then compute residuals $\tilde{\y}_k = \y_k - \A_k \zbar_{(1)}$ for all $k \in \I_1$ and obtain $\S(1)$, $\U_{(1)}$, and $\B_{(1)}$ using function AGM-LpS followed by function AGM-LpS-INIT in Algorithm \ref{lplussMRI}.
For subsequent mini-batches $\ell = 2, 3, \ldots, q/\alpha$ the mean image $\zbar_{(\ell)}$ is updated by minimizing the sum of squared residuals
\[
\arg\min_{\zbar} \sum_{k \in \I_\ell} \|\y_k - \A_k \zbar\|_2^2
\]
subject to the constraint $\|\zbar - \zbar_{(\ell-1)}\|_2^2 \leq \lambda$, ensuring slow temporal changes. The low-rank subspace $\U_{(\ell)}$ is updated by running AGM-LpS initialized at the previous estimate $\U_{(\ell-1)}$ on the residual measurements $\tilde{\y}_k = \y_k - \A_k \zbar_{(\ell)}$.\\
Low-latency reconstruction:
During low-latency reconstruction, for each time $k \in \I_\ell$, the residual measurement is computed as ${\ty}_k = \y_k - \A_k \zbar_{\ell-1}$. The sparse vector $\s_k$ is then estimated by solving the sparse recovery problem $\min_{\s_k} \| \ty_k - \A_k \s_k \|^2$, and the low-rank coefficient vector is estimated as $\b_k = (\A_k \U_{(\ell-1)})^\dagger (\tilde{\y}_k- \A_k \s_k)$.
The unstructured component $\e_k$ is computed using a few iterations of CGLS on the remaining residual $\tty_k = \ty_k - \A_k(\U\b_k + \s_k)$. Finally, the reconstructed image is obtained as
\[
{\z}_k = \zbar_{(\ell-1)} + \s_k + \U_{(\ell-1)} \bhat_k  + \e_k.
\]
We summarize this in the algorithm provided in the Supplemental Material.
\label{als-init}

\subsection{Advantage of the LpS model over the LR only model}
In the presence of a sparse component, that is, when the goal is to recover a low-rank plus sparse (LpS) matrix, AGM-LR fails because it cannot effectively capture the sparse structure as shown in Fig. \ref{Fig_vm}.
To address this limitation, we developed AGM-LpS, which converges reliably and can accurately recover matrices composed of both low-rank and sparse components.
In dynamic MRI settings, FS-LR-MRI and FS-LpS-MRI achieve similar performance when there is no significant motion. However, in the presence of significant motion — such as in the speech dataset — FS-LpS-MRI outperforms FS-LR-MRI. FS-LpS-MRI is also useful for detecting abnormalities in MRI datasets. Although none of the datasets used in this paper contain abnormalities, if they did, the LpS-based methods would be expected to achieve higher accuracy, whereas the LR-based approach would likely perform only fairly.

\Section{Experiments}

\subsection{Simulated Experiments}
For all simulated experiments, the measurement matrices $\A_k \in \mathbb{R}^{m \times n} , \text{ for } k \in [q]$, were generated as independent and identically distributed (i.i.d.) random Gaussian matrices and normalized by scaling each entry by \( 1/\sqrt{m} \). The sparse matrix $\Sstar \in \mathbb{R}^{n \times q}$ was generated to have exactly $\rho$ non-zero entries per column. The positions of these nonzero entries were selected uniformly at random, and their values were independently chosen from the set $\{-a, +a\}$, where $a \in \mathbb{R}$. The low-rank component $\Lstar \in \mathbb{R}^{n \times q} $ was generated as $ \Lstar= \Ustar \Bstar$, where $ \Ustar \in \mathbb{R}^{n \times r} $ is an orthonormal matrix,  and $\Bstar \in \mathbb{R}^{r \times q}$ contains i.i.d. Gaussian entries. The ground truth matrix is defined as $\Xstar = \Lstar + \Sstar $.
Undersampled measurements were simulated as $\y_k = \A_k \xstar_k, \quad \text{for } k \in [q],$
where $\xstar_k$ is the $k$-th column of $\Xstar$. Reconstruction quality was evaluated using the normalized root mean square error (NRMSE), defined as $\frac{\|\Xstar - \X\|_F}{\|\Xstar\|_F}$, where $\X$ is the reconstructed matrix.                                                                                                                                                                                                                                                                                                                                                                                                                                                                                                                                                                                                                                                                                                                                                                                                                                                                                                                                                                                                                                                                                                                                                                                                                                                                                                                                                                                                                                                                                                                                                                                                                                                                                                                                                                                                                                                                                                                                                                                                                                                                                                                                                                                                                                                                                                                                                                                                                                                                                                                                                                                                                                                                                                                                                                                                                                                                                                                                                                                                                                                                                                                                                                                                                                                                                                                                                                                                                                                                                                                                                                                                                                                                                                                                                                                                                                                                                                                                                                                                                         
%
%
\subsubsection{Experiment 1}
\label{exp1}
   \begin{table*}[h]
\begin{center}
\footnotesize
\resizebox{0.7\linewidth}{!}{
\begin{tabular}{|l|l|l|l|l|l|l|l|l|l|}
\hline
$\|\S\|_\infty$ & AGM-LpS-basic  &AltGDmin-LpS   \cite{asilomar} & AGM-LR-basic  \cite{lrpr_gdmin_mri_jp} & GD  \cite{zhang18c} \\
&(proposed)  &($\L$ first) & & \\
\hline
10&   0.0302 (0.23)  &   0.3579 (0.23) &   1.0096 (0.01) &   0.1637 (0.04) \\
\hline
100&   0.0030 (0.23) &   0.3457 (0.24) &   1.0098 (0.01) &   0.0879 (0.04) \\
 \hline

\end{tabular}
}

\caption{\sl{\textbf{Error (Initialization time in seconds)}. The error we report is the normalized RMSE. Our proposed method generally provides better initialization compared to other methods.
 }}
 \label{error_time_init}
\end{center}
\end{table*}
This experiment evaluates the effectiveness of our proposed spectral initialization strategy (see Sec \ref{Uinit}) by comparing it with three existing low-rank plus sparse (LpS) initialization techniques. The other methods we compare against are as follows:
(1) AGM-LpS (L-first) \cite{asilomar}: This approach estimates the subspace $\U$ by assuming $\S = 0$, and then computes $\b_k = (\A_k\U)^\dagger \y_k$ for each frame $k$. The sparse vector $\s_k$ is estimated by solving $\min_{\s_k} \|(\y_k - \A_k\U\b_k) - \A_k\s_k\|^2$. This approach performs poorly when $\Sstar$ has large entries. 
(2) AGM-LR \cite{lrpr_gdmin_mri_jp}: This is a baseline that sets $\S = 0$ and initializes each column of $\L$ using back-projection, i.e., $\A_k^\top \y_k$.
(3) GD \cite{zhang18c}: This method uses a few steps of alternating gradient descent to jointly initialize both $\L$ and $\S$.

We conduct this experiment on an undersampled setting with $n = 100$, $m = 60$, $q = 100$, $r = 2$, and $\rho = 2$, and run 100 Monte Carlo trials. We test three different sparse magnitudes: $\|\S\|_\infty \in \{1, 10, 100\}$. Results are shown in Table~\ref{error_time_init}, which reports NRMSE and runtime in seconds.

The proposed AGM-LpS-basic (Proposed) method consistently achieves the best or second-best performance, especially when the sparse entries are large. In contrast, AGM-LpS (L-first) performs slightly better for very small sparse magnitudes, but fails in other cases. Since AGM-LpS-basic (Proposed) makes no assumptions about the magnitude of $\S$, it is more reliable. 

\subsubsection{Experiment 2}
\begin{figure}
\centering
         \includegraphics[width= 9 cm]{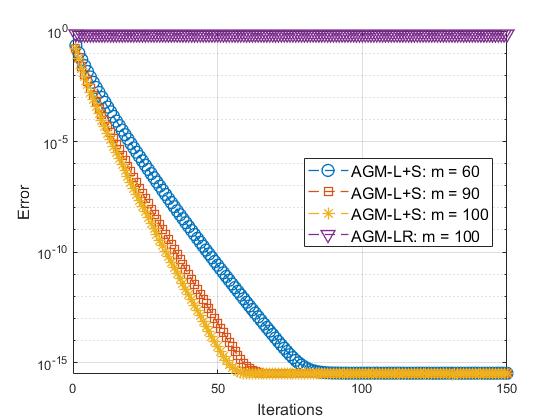}

     \caption{\sl\small{Plot of NRMSE for AGM-LpS-basic with different values of $m$, and in comparison with AGM-LR-basic, versus iteration. Results are averaged over 100 Monte Carlo runs.
}}
\label{Fig_vm}
\end{figure}
In Fig.\ref{Fig_vm}, we study the convergence behavior of the proposed AGM-LpS-basic algorithm under varying values of $m$, and also compare it with AGM-LR. We define convergence as achieving a NRMSE below $10^{-14}$. We use the same parameter settings as in Section \ref{exp1}, with the non-zero entries of the sparse component chosen from the set $\{-1, +1\}$. We vary $m$ across three values: $0.6n$, $0.9n$, and $n$. As shown in Fig.\ref{Fig_vm}, AGM-LR does not converge in the presence of sparse components.

\subsection{MRI Experiments}

\begin{table*}
\begin{center}
\footnotesize
\resizebox{0.8\linewidth}{!}{
\begin{tabular}{|l|l|l|l|l|l|}
\hline
Dataset(Radial) &AGM-LpS &  LpS-Lin  \cite{lin_fessler} & LpS-Otazo \cite{otazo2015low} &kt-SLR \cite{lingala2011accelerated} &   varMRI  \cite{varMRI,varMRI_0}\\
&(proposed) & &  && (DL)\\
\hline
CardPerf (R8)&  \textbf{0.0218} (8.26) & 0.0292 (5.82)& \textbf{0.0110} (18.23)&0.6398 (442.26) & 0.2821 (9.25)\\
& & & &&\\
 \hline
 \hline
CardCine (R6)& \textbf{0.0060} (22.01)&0.0069 (16.40) & \textbf{0.0054} (26.99) & 0.1377 (666.48) & 0.0835 (13.83)\\
& & & &&\\
 \hline
 \hline
CardOCMR16 (4)& \textbf{0.0093} (7.52)& 0.0515 (3.36)& \textbf{0.0293} (11.89)  &0.0362 (264.59) & 0.4837 (4.65)\\
& & & &&\\
 \hline
CardOCMR16 (8)&  \textbf{0.0032} (5.56)& 0.0101 (3.40)&0.0064 (10.31)  &\textbf{0.0045} (265.28) & 0.3932 (6.04)\\
& & & &&\\
 \hline
CardOCMR16 (16)&  \textbf{0.0015} (4.95)&0.0030 (2.90)&0.0035 (5.29) &\textbf{0.0015} (180.99) & 0.2160 (6.20)\\
& & & &&\\
 \hline
 \hline
CardOCMR19 (4)&   \textbf{0.0098} (8.46) &0.0698 (4.82)&0.0251 (19.60) &\textbf{0.0216} (429.31) & 0.1786 (8.64)\\
& & & &&\\
 \hline
CardOCMR19 (8)& \textbf{0.0054} (6.98)& 0.0149 (5.47) &0.0092 (14.05)  &\textbf{0.0043} (438.78) & 0.2101 (5.94)\\
& & & &&\\
 \hline
CardOCMR19 (16)&   \textbf{0.0033} (5.85)&0.0044 (4.85) &0.0052 (7.41) &\textbf{0.0020} (281.20) & 0.1734 (6.80)\\
& & & &&\\
 \hline
 \hline
UnCardPerf (4)&  \textbf{0.0706} (91.93)&0.1424 (48.76)  &0.0910 (194.09)  &\textbf{0.0894} (4232.12) & 0.0733 (66.61)\\
& & & &&\\
 \hline
UnCardPerf (8)&  \textbf{0.0462} (86.40)&0.0632 (48.23)  &0.0591 (122.56) &\textbf{0.0442} (3574.55) & 0.0639 (71.95)\\
& & & &&\\
 \hline
UnCardPerf (16)&  \textbf{0.0293} (70.54)& 0.0329 (48.11) &0.0370 (89.47) &\textbf{0.0206} (2898.11) & 0.0531 (77.38)\\
& & & &&\\
 \hline
 \hline
Speech (4)&  \textbf{0.1349} (142.17)&0.2545 (285.88) &0.1991 (524.47) &\textbf{0.1543} (5189.96) & 0.7990 (746.21)\\
& & & &&\\
 \hline
Speech (8)&  \textbf{0.0969} (133.42)&0.1284 (287.94)& 0.1107 (482.62)  & \textbf{0.0593} (5185.24) & 0.7513 (746.30)\\
& & & &&\\
 \hline
Speech (16)& 0.0576 (150.34)&  0.0557 (288.42)&\textbf{0.0550} (414.29) &\textbf{0.0203} (5214.47) &1.000 (745.47)\\
& & & &&\\
\hline
\hline
av-Err (av-Time) &    \textbf{0.0354} (53.17) &0.0619 (75.31) & \textbf{0.0462} (138.66) &0.0883
 (2090.0) & 0.3401 (179.66)\\
& & & & &\\
\hline
\end{tabular}
}

\caption{\sl{The table format is \textbf{Error (Reconstruction time in seconds)}. The values in parentheses in the first column denote the acceleration factor for Cartesian undersampling and the number of radial spokes for pseudo-radial sampling.  All algorithms were implemented in MATLAB, except for the varMRI deep learning method, which was implemented in Python and run on a GPU (A100). The last row reports the average error and average reconstruction time (in seconds) across all 14 datasets. UnCardPerf stands for free-breathing ungated cardiac perfusion. We also implemented the proposed AGM-LpS method in Python, which achieved an average error (average time) of 0.0354 (16.87). }}


\label{error_batch}
\end{center}
\end{table*}
\begin{table*}
\begin{center}
\footnotesize
\resizebox{1\linewidth}{!}{
\begin{tabular}{|l|l|l|l|l|l|l|}
\hline
Dataset(Radial) & AGM-LpS-FS (modi) &AGM-LpS-FS& KeyHole  \cite{keyhole}  & View-sharing \cite{vs_spiral} & OnAIR \cite{onair} & AGM-MRI-Online \cite{lrpr_gdmin_mri_jp}\\
  & (proposed)& (proposed)   &imaging&& &\\

 \hline
UnCardPerf(4)&  \textbf{ 0.0903} (82.60)& \textbf{ 0.0853} (78.66) &0.2473 (2.06)  &0.5630 (17.73) &0.9540 (6787.18)&\textbf{ 0.1586} (70.05) \\
& & & & & &\\
 \hline
UnCardPerf(8) &  \textbf{ 0.0546} (73.22)&\textbf{ 0.0546} (57.20) & 0.1991 (2.04)  &0.5000 (16.86) & 0.8371 (4767.40)& \textbf{ 0.1271} (36.61) \\
& & & & & &\\
 \hline
UnCardPerf(16)&  \textbf{ 0.0319} (76.96)& \textbf{ 0.0304} (52.84)&  0.1568 (2.07) & 0.3315 (16.28) &0.6011 (4725.84) & \textbf{ 0.0958} (27.35) \\
& & & & & &\\
 \hline
 \hline
Speech(4) &\textbf{ 0.1376} (120.78) & \textbf{ 0.1273} (106.98) &  0.4939 (3.71) &   0.4107 (202.23)& 0.7727 (28233.29) & \textbf{ 0.2977} (37.70) \\
& & & & & &\\
 \hline
Speech(8)&  \textbf{ 0.1001} (116.91)& \textbf{ 0.0920} (98.67) & 0.3981 (3.59) &0.3229 (203.35) &0.5883 (25643.50) & \textbf{ 0.2292} (36.87)\\
& & & & & & \\
 \hline
Speech(16)&    \textbf{ 0.0625} (122.53) &\textbf{ 0.0574} (91.04) &0.3151 (3.59) & 0.2162 (204.68)&0.4881 (28653.84) & \textbf{ 0.1771} (37.45) \\
& & & & & & \\
\hline
\hline
av-Err (av-Time)  & \textbf{ 0.0795} (98.83) &\textbf{ 0.0745} (80.89)& 0.3017 (2.84) & 0.3907 (110.18) & 0.7069 (16469.10) & \textbf{ 0.1809} (41.00)\\
& & & & & &\\
\hline
\end{tabular}
}

\caption{\sl{The table format is \textbf{Error (Reconstruction time in seconds)}. We compare our proposed algorithms, AGM-LpS-FS(modi) and AGM-LpS-FS, with other real-time methods. We used $\alpha=32$ for both our proposed algorithms. All algorithms were implemented in MATLAB. The last row reports the average error and average reconstruction time (in seconds) over all 7 rows. The AGM-MRI-Online results shown here are copied from \cite{lrpr_gdmin_mri_jp}.}} 
\label{error_rt}
\end{center}
\end{table*}
Our code will soon be available at \url{https://github.com/Silpa1}.
The algorithm parameters for AGM-LR were initially tuned using single-coil speech, cardiac, brain, and PINCAT datasets, which were retrospectively sampled in a pseudo-radial manner. It was then evaluated on their multicoil versions, as well as on many other retrospective and prospective datasets sampled using many different sampling patterns 
-- radial, spiral, Cartesian, and pseudo-radial. We then used the same parameters for AGM-LpS, AGM-LpS-FS(modi), and AGM-LpS-FS. 

\subsubsection{Comparison of batch methods}
\label{pros_datasets}
The error metric reported in this section is the normalized scale-invariant mean squared error  (N-S-MSE), defined as: $Error=  (\sum_{k=1}^q \dist^2 (\xstar_k,\hat\x_k) ) / \|\Xstar\|_F^2$, where the scale-invariant distance is $\dist^2 (\xstar,\hat\x) = \| \xstar - \hat\x \frac{\hat\x^\top \xstar}{\|\xhat\|^2}  \|^2$, where ``scale'' is a complex number. We report each result in the format Error (Reconstruction Time in seconds).  

For Table \ref{error_batch}, we used a total of 14 datasets:  12 datasets across four different applications, that were retrospectively pseudo-radially undersampled  with  4, 8 and 16 radial lines  --  free-breathing ungated cardiac perfusion (UnCardPerf), a long but low-resolution speech sequence (Speech),  two cardiac cine datasets from the OCMR database (CardOCMR16, CardOCMR19) \cite{chen2020ocmr}. We employ pseudo-radial sampling to simulate two distinct under-sampling scenarios: (a) 3D Cartesian under-sampling, where phase encoding in the ky - kz plane follows a radial ordering pattern \cite{cheng_pr}, \cite{han_pr}; and (b) true 2D radial under-sampling, in which radial samples are pre-interpolated onto a Cartesian grid using a gridding approach such as GROG \cite{benkert_pr} prior to iterative reconstruction. Additionally,  we included 2 prospectively undersampled datasets from  \cite{lin_fessler}, where Cartesian variable density random undersampling was applied with reduction factors (R); R=8 for cardiac perfusion (CardPerf-R8) and R=6 for cardiac cine (CardCine-R6). Since fully sampled images were unavailable for these two prospectively sampled datasets, we used the converged reconstructions from \cite{lin_fessler} as reference images for error computation and visual comparisons. Here, the converged results  refer to the average of the final reconstructed images from FISTA and POGM after 100 iterations, as  in \cite{lin_fessler}. All datasets were multi-coil.
Image sequence sizes: CardPerf-R8  ($n_x =128, n_y=128, q=40$), CardCine-R6 ($n_x =256, n_y=256, q=24$), CardOCMR16 ($n_x =192,  n_y=150, q=15$), CardOCMR19 ($n_x =192,  n_y=144, q=25$), UnCardPerf ($n_x =288, n_y=108, q=200$) and Speech ($n_x =68, n_y=68, q=2048$). Dataset details are given in the Supplemental Material.  

\begin{figure*}[h]
\centering
         \includegraphics[width= 16 cm]{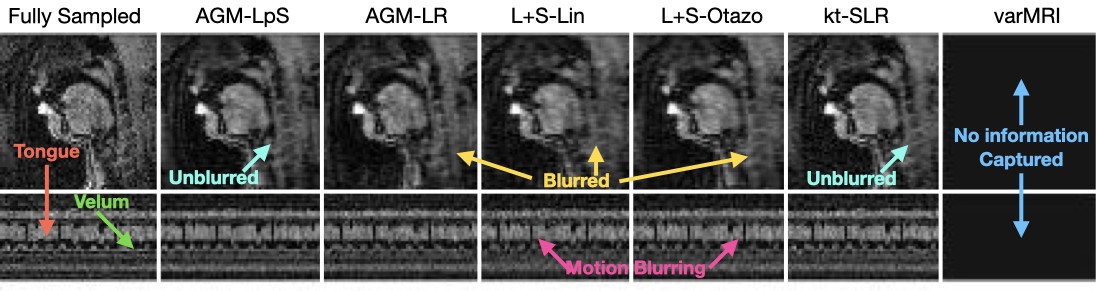}

     \caption{\sl\small{Retrospective, Cartesian (R6), Speech: The first row shows the $500$-th frame of the fully sampled image and its reconstructions using  AGM-LpS (proposed),  AGM-LR \cite{lrpr_gdmin_mri_jp}, LpS-Lin \cite{lin_fessler}, LpS-Otazo \cite{otazo2015low} and k-t-SLR \cite{lingala2013accelerating}. The second row shows the time-profile image to illustrate the motion of the tongue and velum. For simplicity, we display a cut through 100 frames. Compared to AGM-LR, AGM-LpS better captures the motion at the tip of the tongue. While kt-SLR performs well on retrospectively sampled data, it performs poorly on prospectively sampled data (see Fig.\ref{batch_cardperf}). Additionally, kt-SLR requires significantly longer reconstruction time (see Table \ref{error_batch}). The reason VarMRI produces a black image for this dataset is further explained in Section~\ref{pros_datasets}. 
}}
\vsm
\label{batch_speech}
\end{figure*}
\begin{figure*}[h]
\centering
         \includegraphics[width= 12 cm]{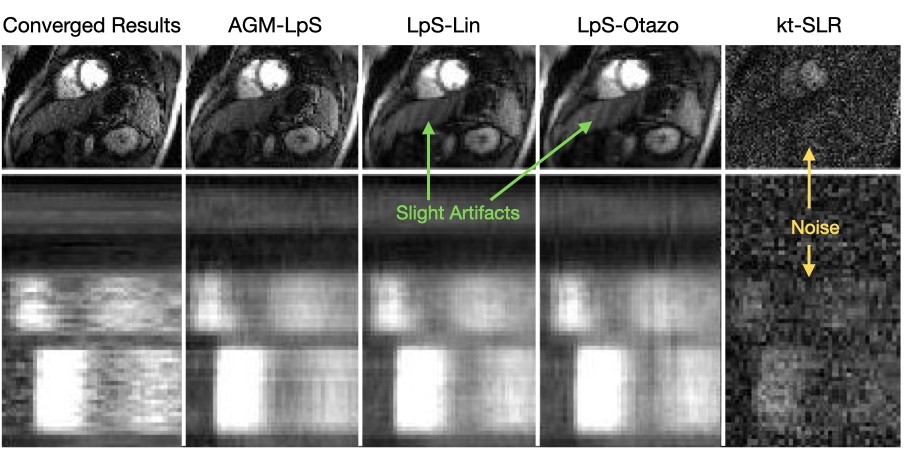}

     \caption{\sl\small{Prospective, Cartesian (reduction factor 8), Cardiac: The first row shows the $10$-th frame of the converged results from \cite{lin_fessler} and its reconstructions using AGM-LpS and with batch methods. Here, the converged results  refer to the average of the final reconstructed images from FISTA and POGM after 100 iterations. The second row shows the time profile image. kt-SLR does not perform well in the presence of noise. AGM-LpS gives better visual results compared to other methods.
}}
\vsm
\label{batch_cardperf}
\end{figure*}
\begin{figure*}[h]
     \centering

       \begin{subfigure}[b]{0.25\textwidth}

         \includegraphics[width= \textwidth]{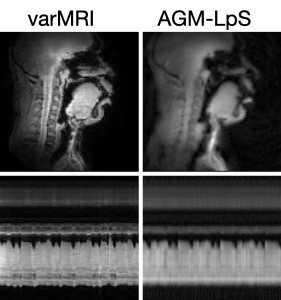}
         \caption{Prospective, Speech, spiral}
         \label{pr_speech_varmri}
          \end{subfigure} \hspace{2 cm}
     \begin{subfigure}[b]{0.36\textwidth}
         \includegraphics[width=\textwidth]{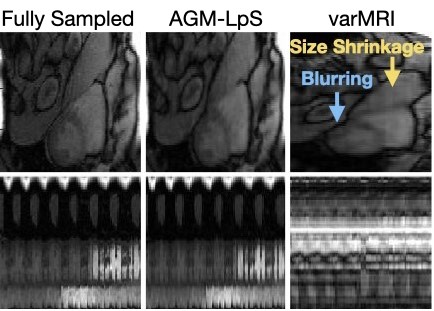}

         \caption{Retrospective, UnCardPerf, pseudo-radial}
         \label{retro_perf}
     \end{subfigure}
          \caption{\sl\small{We demonstrate that varMRI performs well on the dataset it was specifically designed for but fails to generalize to other datasets. In Figs. \ref{retro_perf}–\ref{pr_speech_varmri}, Row 1 shows the reconstructed images, while Row 2 shows the corresponding time profile images. In Fig. \ref{pr_speech_varmri}, we compare recons from varMRI and AGM-LpS. VarMRI gives high image quality recons, as it is specifically designed for this dataset. AGM-LpS(modi2) achieves performance comparable to that of VarMRI. AGM-LpS (modi2) refers to the variant where the view-sharing kspace is used as the input.
In  Fig. \ref{retro_perf}, we compare AGM-LpS with fully sampled recons and varMRI. AGM-LpS produces results comparable to the fully sampled recons and outperforms varMRI.
}}
\label{varMRI}
\vsm
\vsm
\end{figure*}

\begin{figure*}[h]
     \centering
     \begin{subfigure}[b]{0.65\textwidth}
         \centering
         \includegraphics[width=\textwidth]{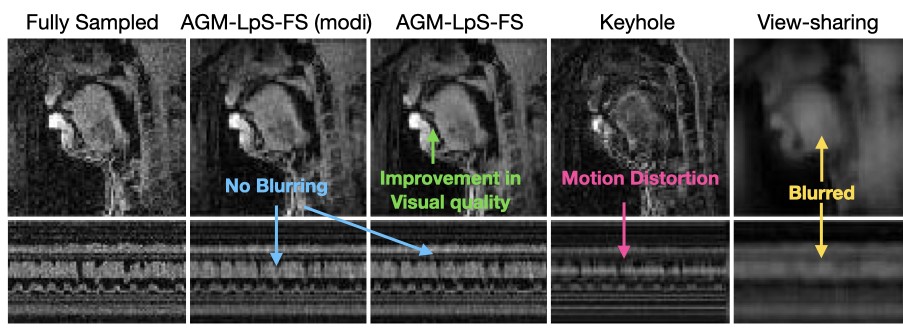}
          \caption{\makebox[\textwidth][l]{\sl{\footnotesize{   Speech, Pseudo-radial (4 radial lines)}}}}
         \label{speech_retro_real}
     \end{subfigure} \hspace{0.8 cm}
     \begin{subfigure}[b]{0.65\textwidth}
         \centering
         \includegraphics[width=\textwidth]{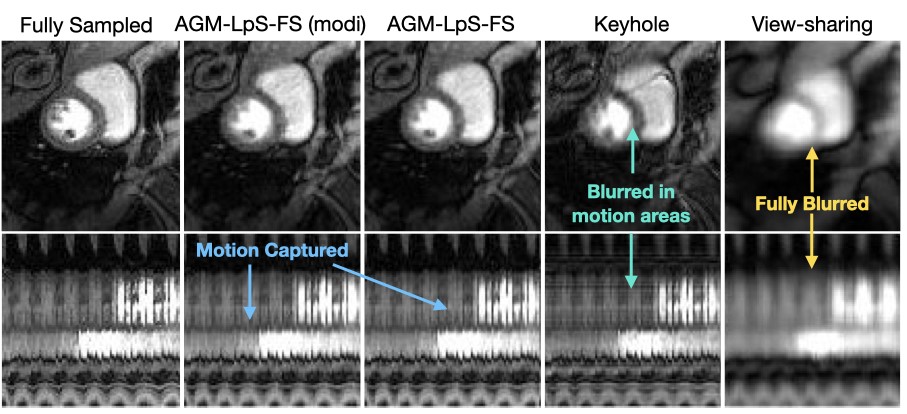}
         \caption{\makebox[\textwidth][l]{\sl{\footnotesize{  Free-breathing ungated cardiac perfusion (UnCardPerf), Pseudo-radial (16 radial lines) }}}}
         \label{card_retro_real}
     \end{subfigure}
\caption{\sl\small{Retrospective: Both figures show fully sampled images and their reconstructions using AGM-LpS-FS(modi) (proposed), AGM-LpS-FS (proposed), keyhole imaging \cite{keyhole}, and view-sharing \cite{vs_spiral}. Row 1 of Fig. \ref{speech_retro_real} and Fig. \ref{card_retro_real} shows the 1470-th and 76-th frames, respectively, while Row 2 displays the corresponding time profile images. AGM-LpS-FS captures localized motion, such as the tip-of-the-tongue movement, more effectively than AGM-LpS-FS (modi), as seen in Fig. \ref{speech_retro_real}. For both datasets, our proposed methods provide improved image quality compared to keyhole imaging and view-sharing.}} 
\label{real}
\vsm \vsm
\end{figure*}

In Table \ref{error_batch}, AGM-LpS  (proposed) is compared with  LpS-Lin \cite{lin_fessler}, LpS-Otazo \cite{otazo2015low}, kt-SLR \cite{lingala2013accelerating} and VarMRI (an unsupervised DL method) \cite{varMRI,varMRI_0}. The last row reports the Average Error (Average Reconstruction Time), computed as the mean over the 14 previous rows. For LpS-Otazo, and LpS-Lin, we used the authors' provided code: LpS-Lin: \url{https://github.com/JeffFessler/reproduce-l-s-dynamic-mri}, LpS-Otazo: \url{https://cai2r.net/resources/ls-reconstruction-matlab-code/}, kt-SLR: code
was emailed by the author to us, VarMRI: VarMRI (\url{https://github.com/rushdi-rusho/varMRI}). 
The VarMRI code requires that the inverse Fourier transform of the undersampled k-space data (zero-filled at unsampled locations) be of size $168 \times 168$ to be compatible with the CNN architecture. To ensure this, we applied low-pass filtering, padding with zeros for smaller dimensions and cropping for larger ones. Without this preprocessing, dimension mismatches in the loss function cause errors.
The original VarMRI code was designed for non-Cartesian sampled data. We replaced the NUFFT operator with the FFT operator in the author's code, which allows efficient processing of Cartesian sampled data. During training in VarMRI, the loss function $\sum_{k \in \mathcal{I}} \| \mathbf{A}_k^\top \mathbf{A}_k \mathbf{x}_k - \mathbf{A}_k^\top \mathbf{y}_k \|^2$ is used, where $\mathcal{I}$ denotes the minibatch indices. This formulation accelerates training for non-Cartesian sampled data. For Cartesian sampled data, the standard loss function $\sum_{k \in \mathcal{I}} \| \mathbf{A}_k \mathbf{x}_k - \mathbf{y}_k \|^2$ is used, as
suggested by the varMRI authors. From Table \ref{error_batch}, we observe that our
proposed method consistently achieve either the best or second-best reconstruction error while also being the fastest.  Fig.\ref{batch_speech} shows a representative frame from the fully sampled speech dataset and its reconstructions using AGM-LpS and several competing batch methods: AGM-LR \cite{lrpr_gdmin_mri_jp}, LpS-Lin \cite{lin_fessler}, LpS-Otazo \cite{otazo2015low}, kt-SLR \cite{lingala2013accelerating}, and VarMRI \cite{varMRI,varMRI_0}. The dataset was retrospectively undersampled using Cartesian sampling with a reduction factor of 
R=6. Among the methods evaluated, AGM-LpS (proposed) and k-t SLR provide superior reconstruction quality. However, k-t SLR is known to be sensitive to noise, as demonstrated on the prospective CardPerf-R8 dataset in Fig. \ref{batch_cardperf}, where its performance noticeably degrades. In contrast, AGM-LpS maintains high reconstruction quality across all tested datasets.
In Fig. \ref{varMRI}, we evaluate the performance of VarMRI~\cite{varMRI,varMRI_0} on both the dataset it was specifically designed for and on other datasets, to assess its generalizability. VarMRI was originally developed to reconstruct a spirally undersampled speech dataset consisting of k-space data with 335 readout points per spiral interleaf, 3 interleaves per frame, 900 frames, 8 virtual coils, and 10 slices, with an image resolution of $168 \times 168$.
In our experiments, we focus on a single slice. Since VarMRI was specifically designed for this dataset, it performs well on it, as shown in Fig. \ref{pr_speech_varmri}. However, as illustrated in Fig. \ref{batch_speech} and Fig. \ref{retro_perf}, VarMRI does not generalize well to other datasets. In Fig. \ref{batch_speech}, VarMRI returns a black image due to improper parameter tuning for that dataset. In Fig. \ref{retro_perf}, for the UnCardPerf dataset, VarMRI exhibits slight blurring and size shrinkage. The size shrinkage is due to the low-pass filtering, as explained earlier.

\subsubsection{Comparison of low-latency methods}
In Table \ref{error_rt}, we compare our proposed algorithms, AGM-LpS-FS(modi) and AGM-LpS-FS, with low-latency reconstruction techniques such as keyhole imaging \cite{keyhole}, view-sharing \cite{vs_spiral}, OnAIR \cite{onair}, and AGM-MRI-Online \cite{lrpr_gdmin_mri_jp}. Keyhole imaging samples the low spatial frequencies (center of k-space) for each frame while reusing the high-frequency components (outer k-space) from a fully sampled reference frame, assuming that most temporal variations occur in the low-frequency region. To ensure a fair comparison, the number of center k-space samples per frame in keyhole imaging matches that used in our proposed algorithms. View-sharing \cite{vs_spiral} enhances temporal resolution in dynamic imaging by integrating phase-encoding lines from multiple frames. Repeated phase-encoding locations are averaged based on their frequency to ensure reconstruction consistency.
Since the authors' implementations of keyhole imaging and view-sharing were unavailable, we implemented these methods in MATLAB. For OnAIR, we used the author provided code: OnAIR: \url{https://github.com/JeffFessler/onair-lassi}. The available OnAIR code is designed for video reconstruction; we modified it for MRI reconstruction. From Table \ref{error_rt}, it is clear that our proposed algorithms, AGM-LpS-FS(modi) and AGM-LpS-FS, consistently achieve the lower error metrics compared to other methods.

Figs. \ref{speech_retro_real} and \ref{card_retro_real} compare AGM-LpS-FS(modi) and AGM-LpS-FS reconstructions with keyhole imaging and view-sharing for the speech and UnCardPerf datasets, respectively. While keyhole imaging and view-sharing offer faster reconstruction times, they often suffer from poor image quality, particularly in the presence of significant inter-frame motion. In contrast, AGM-LpS-FS(modi) and AGM-LpS-FS produce high-quality reconstructions, with AGM-LpS-FS more effectively capturing localized motion than AGM-LpS-FS(modi), as shown in Fig.\ref{speech_retro_real}.

\begin{figure*}[h]
     \centering
     
     \begin{subfigure}[b]{0.44\textwidth}
         \includegraphics[width=\textwidth]{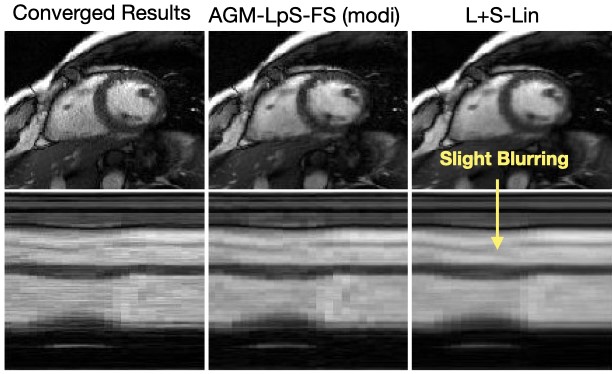}
         \caption{Cardiac Cine, Cartesian (FS-LR-MRI best)}
         \label{pr_cine1}
     \end{subfigure} \hfill
              \begin{subfigure}[b]{0.44\textwidth}
         \includegraphics[width= \textwidth]{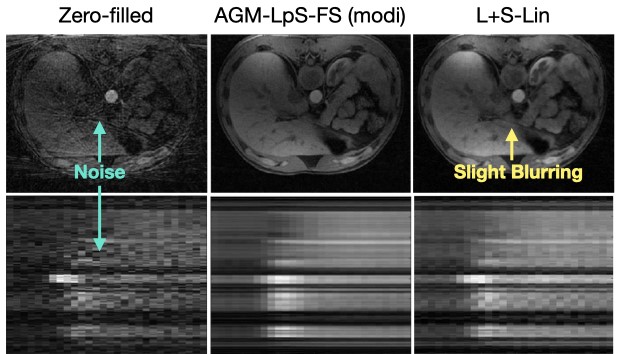}
          \caption{Abdomen, radial, (FS-LR-MRI  best)}
         \label{pr_abdomen1}
     \end{subfigure}\hfill
       
          \caption{\sl\small{Prospective: Figs. \ref{pr_cine1} - \ref{pr_abdomen1} show reconstructions of different MRI applications in Row1 and the corresponding time profile images in Row2. In Fig. \ref{pr_cine1}, AGM-LpS-FS(modi) is compared with the converged results from \cite{lin_fessler} and LpS-Lin. Here, the converged results  refer to the average of the final reconstructed images from FISTA and POGM after 100 iterations.
In Fig. \ref{pr_abdomen1}, AGM-LpS-FS(modi) is compared with zero-filled reconstructions and LpS-Lin. This dataset was used in \cite{lin_fessler}.}}
\label{pros_real}

\vspace{-0.5 cm}
\end{figure*}

\begin{figure*}[h]
     \centering
     \begin{subfigure}[b]{0.44\textwidth}
         \includegraphics[width=\textwidth]{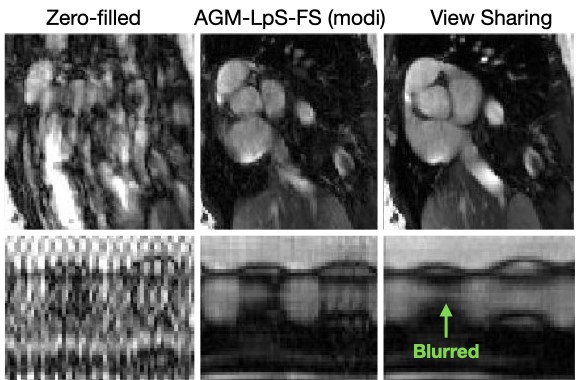}
         \caption{OCMR prospective dataset:$1$, Cartesian (AGM-LpS-FS(modi) best)}
         \label{pr_ocmr1}
     \end{subfigure} \hfill
     \begin{subfigure}[b]{0.44\textwidth}
         \includegraphics[width=\textwidth]{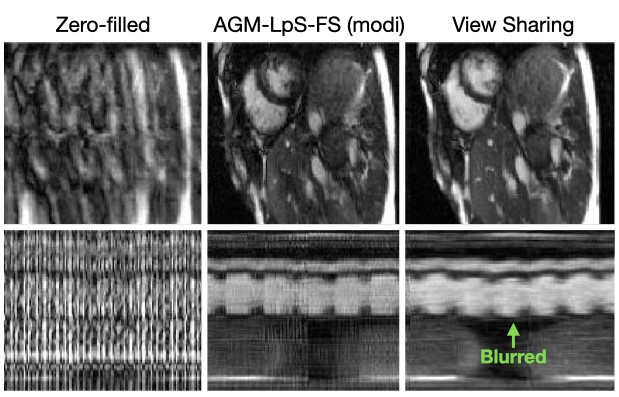}

         \caption{OCMR prospective dataset:$2$, Cartesian (AGM-LpS-FS(modi) best)}
         \label{pr_ocmr2}
     \end{subfigure} 
          \caption{\sl\small{Prospective: Fig. \ref{pr_ocmr1} - \ref{pr_ocmr2} show reconstructions of different MRI applications using AGM-LpS-FS(modi) and other methods in Row 1, while Row 2 shows the corresponding time profile images.
In Fig. \ref{pr_ocmr1} - \ref{pr_ocmr2}, we compare AGM-LpS-FS(modi) with zero-filled and view-sharing. 
}}
\label{ocmr_real}
\vsm
\vsm
\end{figure*}

We show results on four prospectively undersampled datasets: three Cartesian cardiac datasets and one radial abdomen dataset. Note that keyhole imaging cannot be applied to these datasets, as it requires fully sampled k-space data as a reference frame, which is not available in these prospectively acquired datasets.

Our first dataset is CardCine-R6, as described in Sec \ref{pros_datasets}. Fig. \ref{pr_cine1} shows that AGM-LpS-FS(modi) provides good reconstructions for the CardCine-R6 dataset, while LpS-Lin reconstructions exhibit slight blurring in the temporal domain. 
The second dataset is a radially undersampled dynamic contrast-enhanced (DCE) abdomen dataset from \cite{lin_fessler}. To handle non-Cartesian sampling, we incorporate the non-uniform Fast Fourier Transform (NUFFT) function from \cite{lin_fessler} instead of the standard FFT and include density compensation. The k-space data for the DCE abdomen dataset consists of 384 readout points, 21 radial spokes per frame (with golden-angle-based angular increments), 28 time frames, and 7 virtual coils after PCA-based coil compression. The 21 spokes per frame result in a total of 588 spokes ($21 \times 28$), which are distributed across 28 time frames such that each frame is formed by a set of 21 consecutive spokes. As shown in Fig. \ref{pr_abdomen1}, AGM-LpS-FS(modi) and LpS-Lin reconstructions effectively capture contrast uptake dynamics in the liver. While AGM-LpS-FS(modi) clearly resolves other blood vessels, LpS-Lin exhibits slight blurring.
Our next two datasets are Cartesian undersampled cardiac cine prospective datasets from the OCMR database. The first dataset has k-space dimensions of 384 readout points, 14 phase-encode lines per time frame, 137 time frames, and 18 coils, while the second dataset has k-space dimensions of 384 readout points, 16 phase-encode lines per time frame, 65 time frames, and 34 coils. As shown in Figs. \ref{pr_ocmr1}-\ref{pr_ocmr2}, AGM-LpS-FS(modi) reconstructions achieve high-quality results for both datasets, whereas view-sharing introduces motion blurring in both cases.

\Section{Discussion and Conclusions}
In this work, we introduces a set of solution approaches, that rely on the LR and LpS models, for fast and generalizable dynamic MRI. Low latency extensions are developed as well. Here low-latency means the reconstruction time taken per frame (average of 0.04 seconds) is less than the acquisition time per frame. Generalizability, which is the ability to produce accurate and fast enough reconstructions for a large class of MRI applications, sampling schemes and rates, without any parameter tuning, is demonstrated through extensive experiments. We show that our proposed methods outperform the state of the art within MRI for both LR and LpS approaches and unsupervised deep learning methods and for low latency approaches. 


\subsubsection{Limitations}
As observed in our results, the proposed reconstruction algorithm does not achieve the best performance across all applications. In some cases, such as Ungated Cardiac Perfusion (UnCardPerf), blurring artifacts are noticeable when it is heavily undersampled. These artifacts can be mitigated by tuning parameters such as step size, maximum number of iterations, the rank of the low-rank component, and by designing more robust loop exit criterion. However, our goal is to show the performance of the algorithm using a single set of parameters.\\
%
 \bibliographystyle{IEEEtran}
\bibliography{./tipnewpfmt_kfcsfullpap,./refs_Silpa,./byz}
\end{document}